\documentclass[default,iicol]{sn-jnl}

\theoremstyle{thmstyleone}%
\theoremstyle{thmstyletwo}%

\theoremstyle{thmstylethree}%

\usepackage[algo2e]{algorithm2e}
\usepackage{algorithm}

\begin{document}

\title[Article Title]{A stochastic agent-based model to evaluate COVID-19 transmission influenced by human mobility}


\author*[1]{\fnm{Kejie} \sur{Chen}}\email{ckj@dlut.edu.cn}

\author[2,3]{\fnm{Xiaomo} \sur{Jiang}}

\author[1]{\fnm{Yanqing} \sur{Li}}

\author[1]{\fnm{Rongxin} \sur{Zhou}}

\affil[1]{\orgdiv{School of Optoelectric Engineering and Instrumental Science}, \orgname{Dalian University of Technology}, \orgaddress{ \city{Dalian}, \postcode{116024}, \country{China}}}

\affil[2]{\orgdiv{Provincial Key Lab of Digital Twin for Industrial Equipment}, \orgaddress{ \city{Dalian}, \postcode{116024}, \country{China}}}

\affil[3]{\orgdiv{School of Energy and Power Engineering}, \orgaddress{ \city{Dalian}, \postcode{116024}, \country{China}}}


\abstract{
The COVID-19 pandemic has created an urgent need for mathematical models that can project epidemic trends and evaluate the effectiveness of mitigation strategies. To forecast the transmission of COVID-19, a major challenge is the accurate assessment of the multi-scale human mobility and how they impact the infection through close contacts. By combining the stochastic agent-based modeling strategy and hierarchical structures of spatial containers corresponding to the notion of places in geography, this study proposes a novel model, Mob-Cov, to study the impact of human traveling behaviour and individual health conditions on the disease outbreak and the probability of zero COVID in the population. Specifically, individuals perform power-law type of local movements within a container and global transport between different-level containers. Frequent long movements inside a small-level container (e.g. a road or a county) and a small population size reduce the local crowdedness of people and the disease infection and transmission. In contrast, travels between large-level containers (e.g. cities and nations) facilitate global disease spread and outbreak. Moreover, dynamic infection and recovery in the population are able to drive the bifurcation of the system to a "zero-COVID" state or a "live with COVID" state, depending on the mobility patterns, population number and health conditions. Reducing total population and local people accumulation as well as restricting global travels help achieve zero-COVID. In summary, the Mob-Cov model considers more realistic human mobility in a wide range of spatial scales, and has been designed with equal emphasis on performance, low simulation cost, accuracy, ease of use and flexibility. It is a useful tool for researchers and politicians to investigate the pandemic dynamics and plan actions against the disease.
}




\maketitle

\section{Introduction}\label{sec1}

COVID-19 is a highly infectious disease transmitted through several ways, including direct physical contact, inhalation of small particles and droplets disseminated by sneezing or coughing, and virus particles or droplets landing on eyes, nose or mouth\cite{1}.  On January 30, 2020, the outbreak of COVID-19 is declared as a Public Health Emergency of International Concern by the World Health Organization (WHO), which has led to the global recession. Until now, the disease is still heavily affecting the human health and economies of most countries\cite{2,3}. The COVID-19 pandemic has created an urgent need for models to analyze and predict the interactions between the infectious disease and human beings.

Models for studying the pandemic spread and the impact of intervention strategies can be broadly divided into two types: partial differential equation (PDE)-based or stochastic differential equation (SDE)-based compartmental models\cite{4,5,6,7,8,9,9extra,9extra2} and agent-based Monte Carlo models\cite{10,11,12,13,14,15,16,17,17extra}. The compartmental models are simple and computationally cost-effective, and thus have been deployed to analyze the spread of COVID-19 in various places, including Wuhan\cite{5}, Germany\cite{6} and Italy\cite{7}. For example, Dehning et al.\cite{6} combined the susceptible-infected-recovered (SIR) model with Bayesian parameter inference to detect change points of disease spreading rate using COVID-19 case numbers measured in Germany. On the other hand, agent-based models consider the individual differences, microscale policies and detailed actions, which can provide a comprehensive understanding of the transmission dynamics  as well as a flexible framework for implementing various intervention rules\cite{10,11,12,13,14,15,16,17,18,19,19extra,19extra2,19extra3}. The agent-based models have been used to study the spread of COVID-19 in Australia\cite{10}, Singapore\cite{11} and France\cite{12} under various interventions\cite{19extra3}. For example, Hoertel et al.\cite{12} developed an agent-based disease transmission model on the social contact network. They examined the impact of lockdown period and post-lockdown measures (e.g. physical distancing, mask wearing) on the cumulative disease incidence and intensive care unit (ICU)-bed occupancy. Meanwhile, several user-friendly, high-performance COVID-19 microsimulation softwares, including the OpenABM-Covid 19\cite{14} and Covasim\cite{15}, have been developed for researchers and public health officials to easily study the COVID-19 dynamics and inform policy decisions.

However, the PDE/SDE-based compartmental models neglect the physical infection process\cite{4,5,6,7,8,9,9extra,9extra2}, which loses accuracy in reflecting real transmission dynamics. Most agent-based models and softwares consider individual contacts using abstract networks and build disease propagation model on these networks. There are three widely-used networks, i.e., microscopic commute networks for work, study and other activities\cite{10,11,13,14,15}, macroscopic commute networks for transport between cities and countries\cite{20,21,22,23}, and social contact networks among family members, friends, workmates, etc.\cite{12}. In addition, some agent-based models consider individual movements in a continuous physical space\cite{16,17,18,19,19extra2}. For example, Cuevas \cite{16} studied how the combination of local random walks and long directed movements of individuals influences COVID-19 transmission. Silva et al.\cite{17} incorporated four types of human behaviors, including walking freely, going home, going to work and going to hospital, into the infection model, where the four walking behaviors are modeled as random walks of different velocities and directions. 

Though the commute and contact networks can partially reflect the characteristics of human mobility, these networks are usually biased by small population samples used to estimate the network topology and weights of edges\cite{20}. Furthermore, the commute networks only reflect travels in a limited number of areas in the same spatial scale, either the microscopic scale containing home, workplace, school and hospital\cite{10,11,13,14,15,19extra2}, or in the macroscopic scale containing cities and countries\cite{21,22,23}. The social contact networks are naturally formed in a long period of time. But the transient contacts between strangers and acquaintances are determined by many factors besides their social relationships\cite{24extra1}. In addition, in the microscopic continuous space, studies have also showed that human mobility patterns do not follow a random walk or a scale-free l\'evy walk\cite{24extra2, 24extra3}.

Recently, Alessandretti et al.\cite{24} have showed that physical space can be represented by nested containers with associated size, satisfying the intuitive conception of space which is hierarchical and characterized by typical scales. Human moving traces modeled as the combination of commuting within the same containers and transporting between containers of different scales are more realistic. Specifically, they proposed that the physical space of each individual can be represented as a hierarchy of $L$ levels ordered from the smallest to largest. The small level contains individual positions, rooms or houses. A large level contains states or countries. At each level, the physical space is partitioned into compact containers with certain characteristic sizes. The individual travels among the containers at the same or different levels based on the probabilities proportional to the container distance and the level distance. Compared with other state-of-art models\cite{24extra1,24extra2,24extra3,25,28,29,29extra1,29extra2}, the hierarchical container model provides an unbiased performance estimate and a significant better description of human GPS traces evaluated using several index including the distribution of displacements, evolution of radius of gyration and time allocation among locations and entropy. Therefore, modeling COVID-19 infection and propagation in the nested containers can provide more accurate and  insightful results about how human mobility influences disease transmission dynamics. 

\begin{figure*}[!t]
    \centering
    \includegraphics[width=0.9\linewidth]{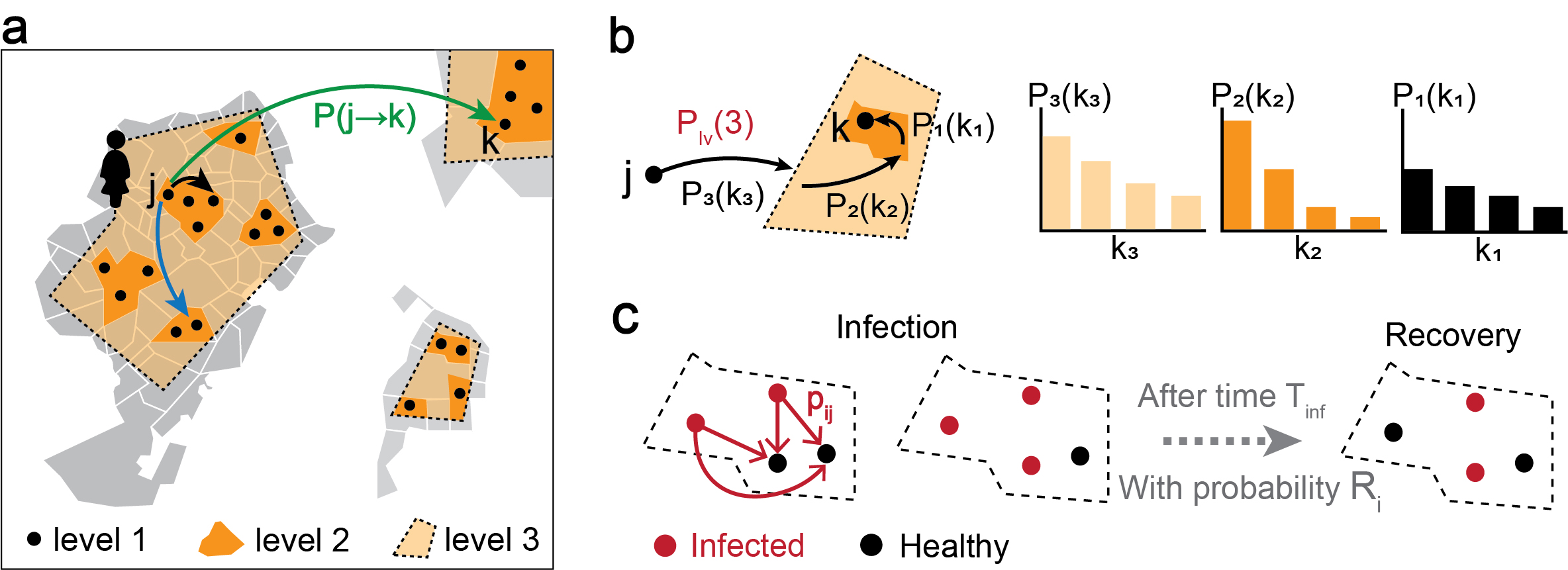}
    \caption{\textbf{Schematics of the Mob-Cov model.} \textbf{(a)} Human mobility on the three-level nested container structures. \textbf{(b)} The probability of traveling from location $j$ to $k$ is determined by two steps: selecting a level (e.g. selecting level 3 by $P_{lv}(3)$) and selecting the nested containers level by level (e.g. $P_3(k_3)$, $P_2(k_2)$, $P_1(k_1)$). \textbf{(c)} The COVID-19 infection and recovery rules.}
\end{figure*}

In this study, we present a stochastic agent-based model, Mob-Cov (mobility influenced COVID-19 transmission model), which considers the hierarchical geographical mobility patterns and the COVID-19 infection and recovery processes. Specifically, in the model, each person is represented by an agent which can move around freely and has an individual-specific health condition represented by the infection and recovery rates. Assuming that 0.5\% of the population is infected initially. As time goes on, people surrounding the infected individuals (i.e. these people are in the same container with the infected ones) also get infected. Sick people travels to other containers, which turns the local disease transmission to a global problem and eventually induces the outbreak crisis. Next, the recovery of sick people after an infection period is included. Instead of having an outbreak, the system is able to achieve a dynamical equilibrium with the coexistence of the healthy and infected people. Furthermore, the influence of population size, health conditions, hierarchical geographical structures and population mobility on the disease transmission dynamics are predicted. Overall, the Mob-Cov model and the results presented in this study provide a novel and comprehensive view of the spread of COVID-19 from a multi-scale human mobility perspective, and can be valuable and helpful in evaluating and designing better interventions and mitigation measures for the pandemic.

\section{The proposed Mob-Cov model}
\subsection{Agent-based model}
To investigate the impact of human mobility on the transmission of COVID-19, we develop a stochastic agent-based Mob-Cov model. The model includes three key components: (1) a synthetic population generated with the hierarchical geographical characteristics (Figure 1a); (2) realistic human travel rules, taking into account the power law behaviors of travelling within a container and transporting between containers (Figure 1b); (3) the infection and recovery with probabilities influenced by individual health conditions (Figure 1c). The first two components about human mobility are discussed in subsections II B and II C. The last component about disease infection and recovery is discussed in the subsection II D. The flow chart (framework) of the Mob-Cov model is shown in Figure 2. 

This model framework includes a flexible individual-based approach that can capture not only the local interactions between individuals and microscopic disease transmission, but also the global spread of the disease influenced by the really long-distance cross-district travels. The framework allows policymakers to define measures at various spatial-temporal scales and evaluate the effective length scale of the interventions.

In the following sections, the three key components of the model and their mathematical foundations are first introduced. The selection of model parameters and the evaluation methods are then presented. Lastly, the pseudo code of the model and the computational procedures are given.

\subsection{Human mobility in the nested containers}
The hidden geographical structure of the population is modeled using nested containers proposed by Alessandretti et al.\cite{24}. Specifically, the physical space is assumed to contain hierarchy of $L$ levels. Level 1 is the smallest and level $L$ is the largest. At any level $l$ ($l=1, 2,...,L$), the space is partitioned into $n_l$ topologically compact containers. For $l<L$, a container is fully included within a single parent container. Thus, each geographical location $k$ can be uniquely identified as a sequence of containers ($k=(k_1, k_2, ..., k_L)$), where the child container at $l$ level $k_l$ is fully included in the mother container at $l+1$ level $k_{l+1}$. For example, Figure 1a shows the schematics of a three-level hierarchical structure. At the largest level (level 3), there are 3 containers, as shown by the light orange polygons. Level 2 has 9 containers, as shown by the orange polygons. And level 3 contains 21 individual positions, represented by the black dots. The location of agent $j$ is defined as $j=(j_1, j_2, j_3)$, where $j_3$ is the ID of the left large polygons in level 3, $j_2$ represents the middle polygon in $j_3$ and $j_1$ is the ID of one black dot in $j_2$.

\begin{figure*}
    \centering
    \includegraphics[width=0.85\linewidth]{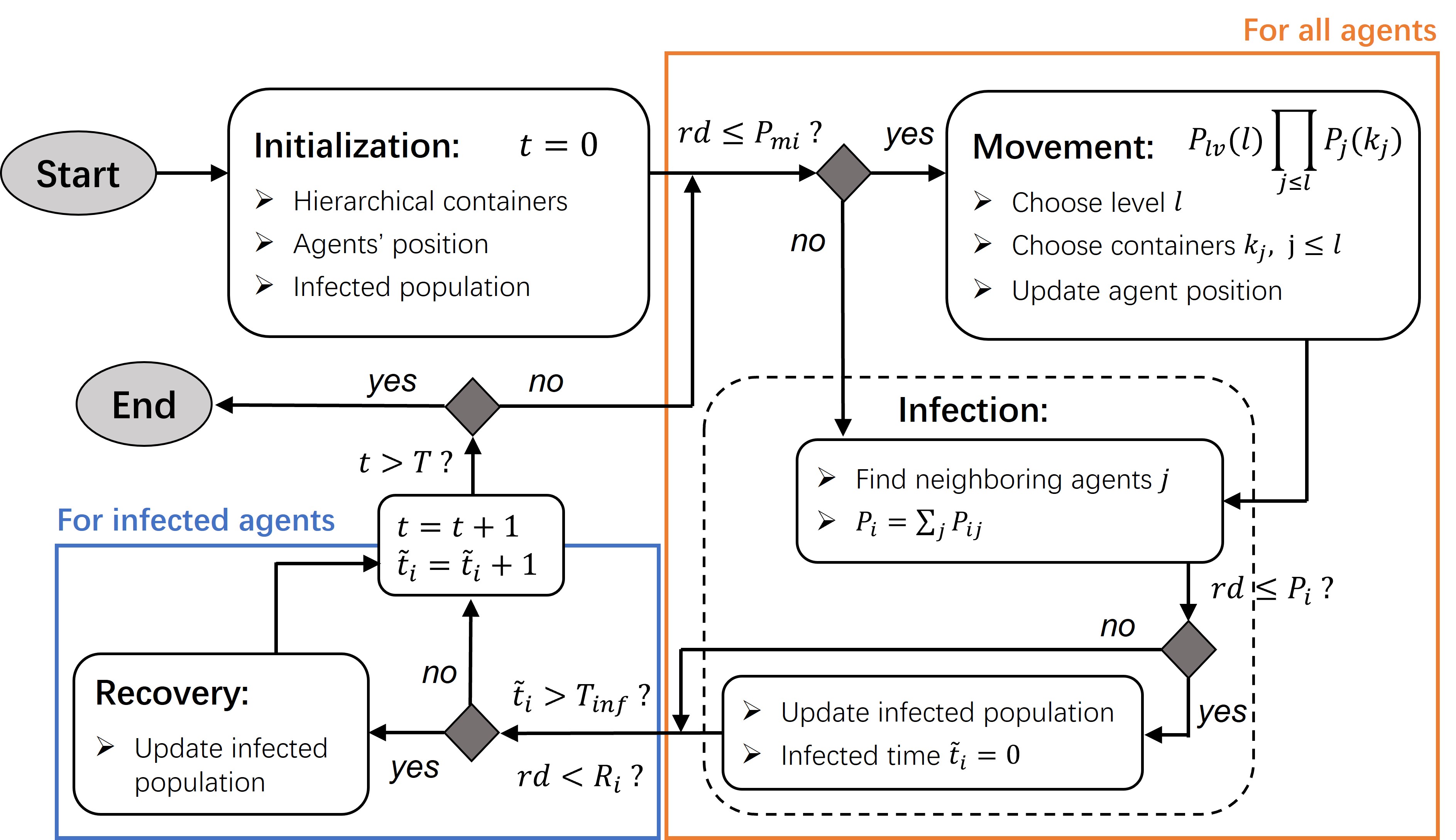}
    \caption{\textbf{The flowchart of the Mob-Cov model.} $rd$ is a random number uniformly distributed between 0 and 1. $t$ is the current simulation time. $\tilde{t}_i$ is the infection time of agent $i$. $P_{mi}$, $P_{lv}(l)$, $P_j(k_j)$, $P_i$, $R_i$ and $T_{inf}$ are model parameters discussed in section II.}
\end{figure*}

Travelling in the nested containers follows a two-stage process: first deciding which level to travel at, second selecting a container at that level, and then a container within the chosen container and so on down to the lowest level. Specifically, suppose an agent is at position $j=(j_1, j_2,..., j_L)$, the probability for the individual to travel to position $k=(k_1, k_2,...,k_L)$ can be expressed as (Figure 1b)
\begin{equation}
    P(j \rightarrow k) = P_{lv}(l)\prod_{i \leq l} P_i(k_i)
\end{equation}
where $P_{lv}(l)$ is the probability of the agent choosing level $l$. The agent stays in the same container if the container level is larger than $l$. In $1, 2, ..., l$ levels, the agent switches to different containers. Thus, $k_{l+1}=j_{l+1}$, $k_{l+2}=j_{l+2}$, ..., $k_{L} = j_{L}$, but $k_{0}\neq j_{0}$, $k_{1} \neq j_{1}$, ..., $k_{l} \neq j_{l}$. $P_{lv}(l)$ follows an exponential distribution with a parameter $d$, which is expressed as
\begin{equation}
    P_{lv}(l) = e^{-d \cdot l},\, \,\,\,\forall \,\,\, l\in [1, L]
\end{equation}
where $l$ is an integer representing the level. Eq. (2) indicates that switching between large-level containers (e.g. between countries) is less frequent than switching between small-level containers (e.g. between houses or streets).

$P_i(k_i)$ is the probability of choose container $k_i$ at level $i$. $P_i(k_i)$ follows a long-tail distribution with parameters $c_0$ and $c_1$, which is expressed as
\begin{equation}
P_i(k) = k^{-(c_0+c_1 \cdot i)} ,\, \,\,\,\forall \,\,\, i\in [1, l]
\end{equation}
where $i$ is the level. $k$ is the container position. Eq. (3) suggests that the travel probability decreases over the distance, especially at a large level $i$. This indicates that, for example, traveling to a distant country is more scarcer than going to a remote road. But compared with the widely-used exponential distribution, the long-tail distribution allows the happening of long-distance movements at all levels.

Eqs. (1)-(3) calculate the probability of switching from location $j$ to location $k$ if the agent takes a movement. We consider the fact that people do not travel at every time moment but need to take a rest or stay at a place for a while. Thus, the probability for an agent to take a movement is modeled with a probability $P_m$, which is uniformly distributed between 0.2 and 0.4. A portion of people with low movement tendency (i.e. low mobility population) is further considered. The ratio of the low mobility people in the population $r_{lm}$ is a model parameter. The probability for the low mobility people to take a movement is only $P_m=0.03$.

\subsection{Infection and recovery}
If two agents $i$ and $j$ ($i$ is infected and $j$ is healthy) are in the same level-2 container (level-1 containers are the individual positions), infection happens with a probability $P_{ij}$ (Figure 1c). If there are more than one infected agents, the infection probability of a healthy agent $j$ is calculated as $P_j = 1-\prod_{i} (1-P_{ij})$, for all infected agents $i$ in the same level-2 container with $j$. For the high risk population, $P_{ij}$ follows a uniform distribution between 0.1 and 0.3. For the low risk population, $P_{ij}=0.02$. The low risk people ratio in the population $r_{li}$ is a model parameter. Note that, $P_{ij}$ is chosen according to previous studies \cite{15, 16}. Kerr et al. \cite{15} set $P_{ij}=0.05$ for household-level transmission. Cuevas \cite{16} set $P_{ij}=0.2$. 

After being infected for a time period $T_{inf}$, the individual $i$ is able to recover to a healthy state with a probability $R_i$ (Figure 1c).

\subsection{Outbreak endpoint}
To evaluate the spread speed of the disease in the population, the outbreak endpoint is calculated. Using the outbreak endpoint instead of the time when all the agents are infected is because there are always few individuals having really low infectious probability $P_{ij}$. The time it takes to get the whole population infected is heavily biased by the last few individuals, which cannot directly and accurately reflect the system property. Therefore, an outbreak endpoint which considers the effects of both the infected population ratio and the infection rate is used in this study\cite{13,30}.

The detection of the outbreak endpoint is similar to find the knee point in system engineering. The knee point usually represents the "right decision point", after which the system properties are no longer significantly influenced by the variation of the variable\cite{30}. Thus, in this study, we follow the previously proposed methods and calculate the maximum point of the cost function. The maximum point is defined as the outbreak endpoint of the system. The cost function is expressed as\cite{13}
\begin{equation}
J(t) = \frac{N_{inf} (t)}{N_p} - \frac{t}{T_{max}}
\end{equation}
where $N_{inf}$ is the number of infected agents at time step (iteration step) $t$, $N_p$ is the total population number, $T_{max}$ is the total time (maximum iteration number), which is defined as the time when 90\% of the population got infected plus 100 extra time steps.

It has been proved that the cost function $J$ can only have one global maximum point, which corresponds to the outbreak endpoint\cite{13}. Therefore, the outbreak time can be calculated as
\begin{equation}
T_{b} = \arg_{t \in [1, T_{max}]} \, max \, J(t)
\end{equation}

Figure 3 shows a simulation example of the infection rate over time and the determination of the outbreak endpoint. Figure 3a shows that the number of newly infected people is the largest during the first 70 time steps. After 70 time steps, the infection rates gradually decrease to values close to zero. As expected, the proportion of infected people keeps increasing over time, but the increasing speed slows down (Figure 3b). By finding the maximum point of the cost function (Figure 3c), the outbreak point can be identified when $t$ is around 70, which corresponds to an infection ratio around 80\%. The results suggest that the first 70 time steps are the most critical period when the number of infected people increases rapidly. After around 80\% of the population gets infected, the infection dynamics slows down. And it can take a really long time (can be longer than 2000 time steps) for the whole population to get infected.

It should also be noted that the outbreak happens when only infection of individuals is included in the model. After considering the recovery of sick people, the system is likely to reach a dynamical equilibrium with both infected and healthy population, instead of reaching the outbreak endpoint. 

\begin{figure*}
    \centering
    \includegraphics[width=0.95\linewidth]{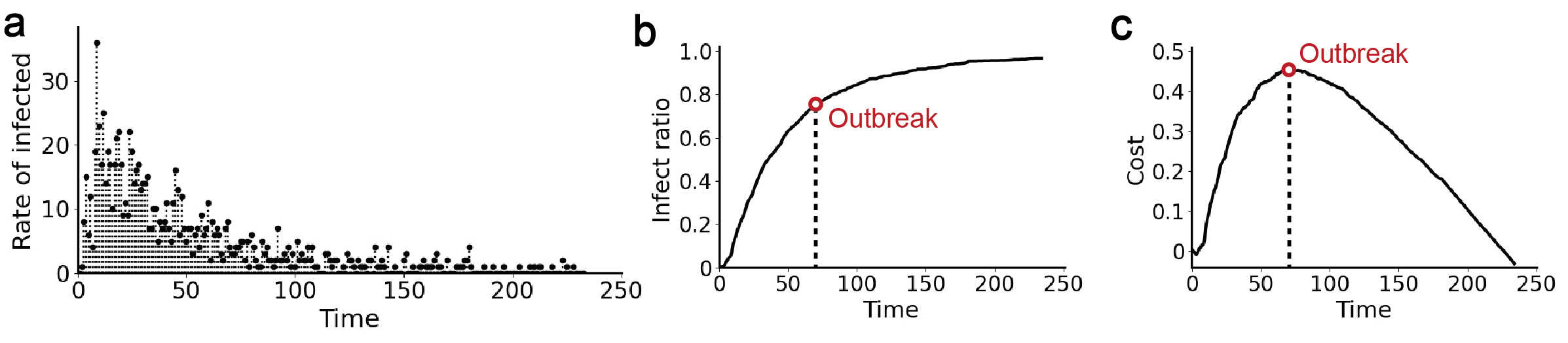}
    \caption{\textbf{The evolution of the infection process} \textbf{(a)} The rate of infection (the number of new infected people) over time. \textbf{(b)} The infection ratio of the population over time. \textbf{(c)} The cost function $J(t)$ over time. The outbreak endpoint is the global maximum point of the cost function, labeled in red.}
\end{figure*}

\subsection{Computational procedure}
The Mob-Cov model is implemented as an iterative scheme. The flowchart is shown in Figure 2 and the pseudo-code is shown below. In the pseudo-code, $L$ is the number of levels. $N_{cont} = (n_1, n_2, ..., n_L)$ is the number of containers in each level. $c_0$, $c_1$ and $d$ are parameters that determine the mobility, as discussed in section II2. $N_p$ is the number of agents. $N_{inf0}$ is the number of initially infected agents. $T_{inf}$ is the infection time period before the agent having a probability to recovery. $r_{lm}$ is the ratio of low mobility population. $r_{li}$ is the ratio of low infection risk population. NestedCont is the nested container structure, which is a nested random dictionary implemented in Python. $P_{mi}$ is the probability of taking a movement for agent $i$. $P_{ij}$ is the probability of agent $j$ getting infected by agent $i$. $R_i$ is the recovery rate of agent $i$. If agents $i$ and $j$ are in the same level-2 container, $j$ is defined as the neighboring agent of $i$.

\floatname{algorithm}{}
\renewcommand{\thealgorithm}{}
\begin{algorithm}[H]  
	\caption{\textbf{Mob-Cov model}}
	\LinesNumbered 
	\textbf{Input:} $L$, $N_{cont}$, $c_0$, $c_1$, $d$, $N_p$, $N_{inf0}$, $T_{inf}$ \vspace{0.2cm}
	
        \textbf{Initialise:} 
        
        NestedCont (nested container structure) 
        
        agent position $\mathbf{X}_i$ ($i=1, 2, .., N_p$)
        
        infected agent IDs

        $P_{mi}$ (mobility) for each agent $i$
        
        $P_{ij}$ (infection) for agents $i$ and $j$ ($i \neq j$)

        $R_i$ (recovery) for each agent $i$ \vspace{0.2cm}
    
	\While{(iter $<$ MaxIteration) or (stop criterion)}{
        \vspace{0.1cm}
		\For{each agent $i$}{
            \vspace{0.1cm}
            \If{(rand $\leq P_{mi}$)}{
            \vspace{0.1cm}
            Update $\mathbf{X}_i$ using Eqs. (1)-(3)
            }}
    	\vspace{0.2cm}

            \For{each infected agent $i$}{
            \vspace{0.1cm}
            \For{each neighboring healthy agent $j$}{
            \vspace{0.1cm}
            \If{(rand $\leq P_{ij}$)}{
            \vspace{0.1cm}
            Update agent $j$ as infected
            }}
            \vspace{0.1cm}
            \If{(Infection time $> T_{inf}$) \& (rand $\leq R_i$)}{
            Update agent $i$ as healthy
            }
            
            }
        }
\vspace{0.2cm}
Post process results and visualization
\end{algorithm}

In the simulation, the maximum iteration number is set as 2000, which is long enough for the system to reach the equilibrium.The number of agents is 1000. The number of initially infected people is 5. For each experiment, 50 simulations are performed, and the mean and variance of the results (including the outbreak time, infected population ratio when outbreak, infected ratio, zero-COVID ratio) are calculated.

\section{Results}
\subsection{The spread dynamics of COVID-19}
Figure 4a shows a simulation result of the population distribution and the spread of COVID-19 disease. In the simulation, the number of levels is $L=5$. Parameters of mobility are $d=1.5$, $c_0=0.8$, $c_1=0.35$. The number of containers in each level is $N_{cont}=(8, 4, 9, 9, 250)$. Containers in level 5 are represented by colorful and transparent polygons. Individuals in different level-5 containers are represented by dots labeled with different colors. Figure 4b is the probability of selecting a container level, $P_{lv}(l)$. The grey bars are simulation results, which fit well with the analytical expression in Eq. (2). Figure 4c is the probability of selecting a container in level 2, $p_2(k)$. The simulation results represented by the black dots fit well to the analytical expression in Eq. (3).

\begin{figure*}[!t]
    \centering
    \includegraphics[width=0.95\linewidth]{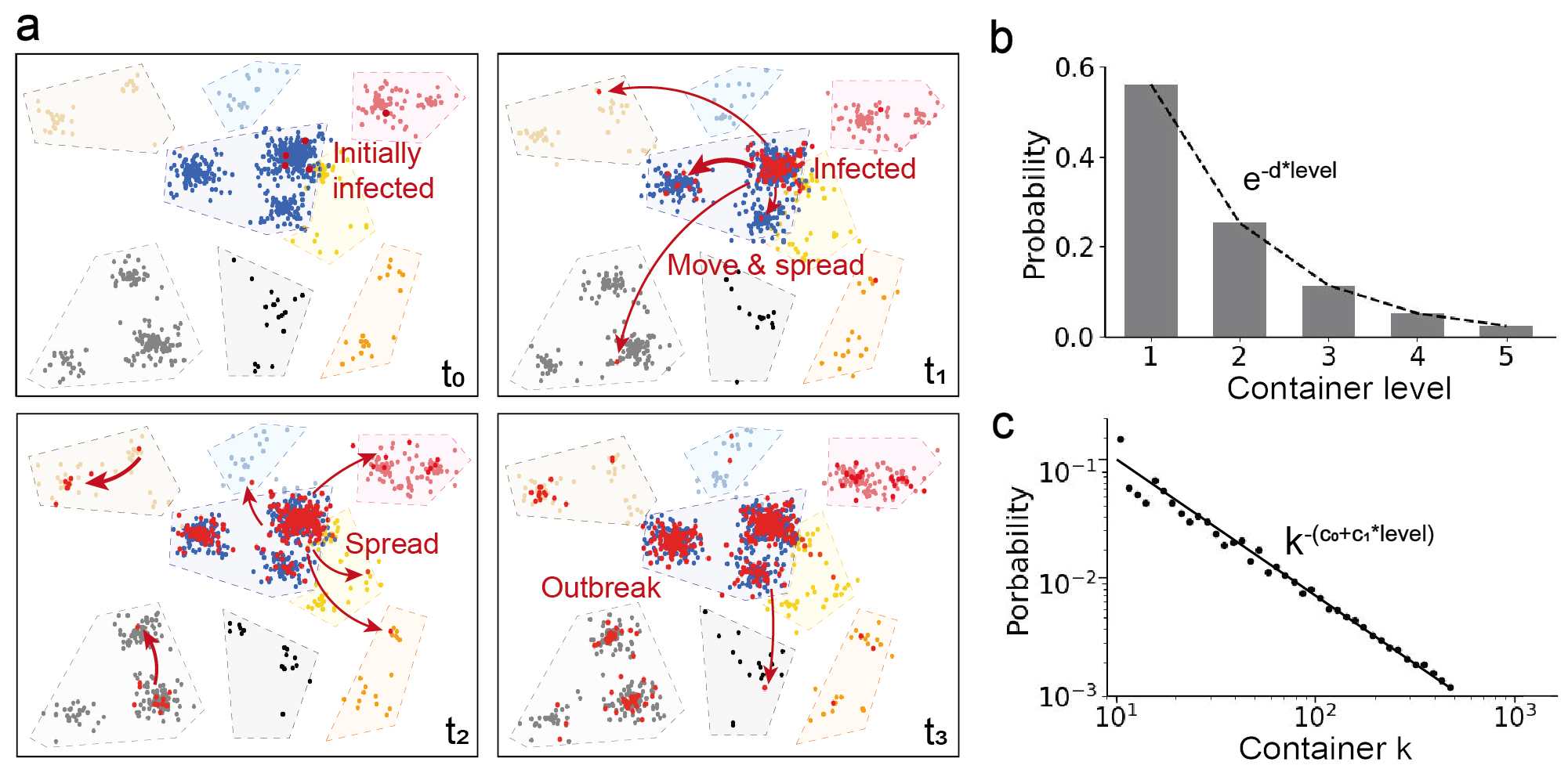}
    \caption{\textbf{Visual results of the population mobility and the spread of the disease.} \textbf{(a)} Distribution of the health and infected individuals at four time steps. In the simulation, $L=5$, and there are 8 containers at level 5, which are represented by the colorful polygons. Individuals in the 8 containers are represented by the colorful dots. Infected individuals are labeled as red. \textbf{(b)} Probability of switching to level $l$ ($l\in [1,5]$). The grey bars are the simulation results. The dotted line is calculated using Eq. (2). \textbf{(c)} Probability of moving to container $k$ at level 2. The black dots are the simulation results, and the black line is calculated using Eq. (3).}
\end{figure*}

At the beginning, only five individuals (four in the blue container and one in the red container) are infected. Due to the high population density in the blue container, the disease quickly spreads out in one cluster at time $t_0$. Many infected people then travels to the nearby cluster in the same blue container, accelerating the local transmission. Meanwhile, the spread of the disease is relatively slow in the red container due to the low population density. Afterwards, one or two infected people travel between the level-5 containers and reach the grey and brown areas, indicating that the disease is transmitted globally. After a further local spread at time $t_2$, a disease outbreak happens at $t_3$.

Overall, the spread of the disease is a multi-scale dynamic process influenced by the probability of switching between levels (Figure 4b) and the movement distance (Figure 4c). The results indicate that the total population density and the initial infected people number does not need to be really large. When the disease starts from a local population cluster with a high people density (e.g. a crowded market), a small probability of traveling between large-level containers (e.g. traveling cross cities or countries) can lead to a rapid global spread and outbreak. The local-to-global spread of the disease heavily relies on the long movements between containers. Thus, accurately modeling both the short walks and long sojourns using the hierarchical container structure in the Mob-Cov model ensures the precise prediction of the disease dynamics. 

\begin{figure*}[!t]
    \centering
    \includegraphics[width=1.0\linewidth]{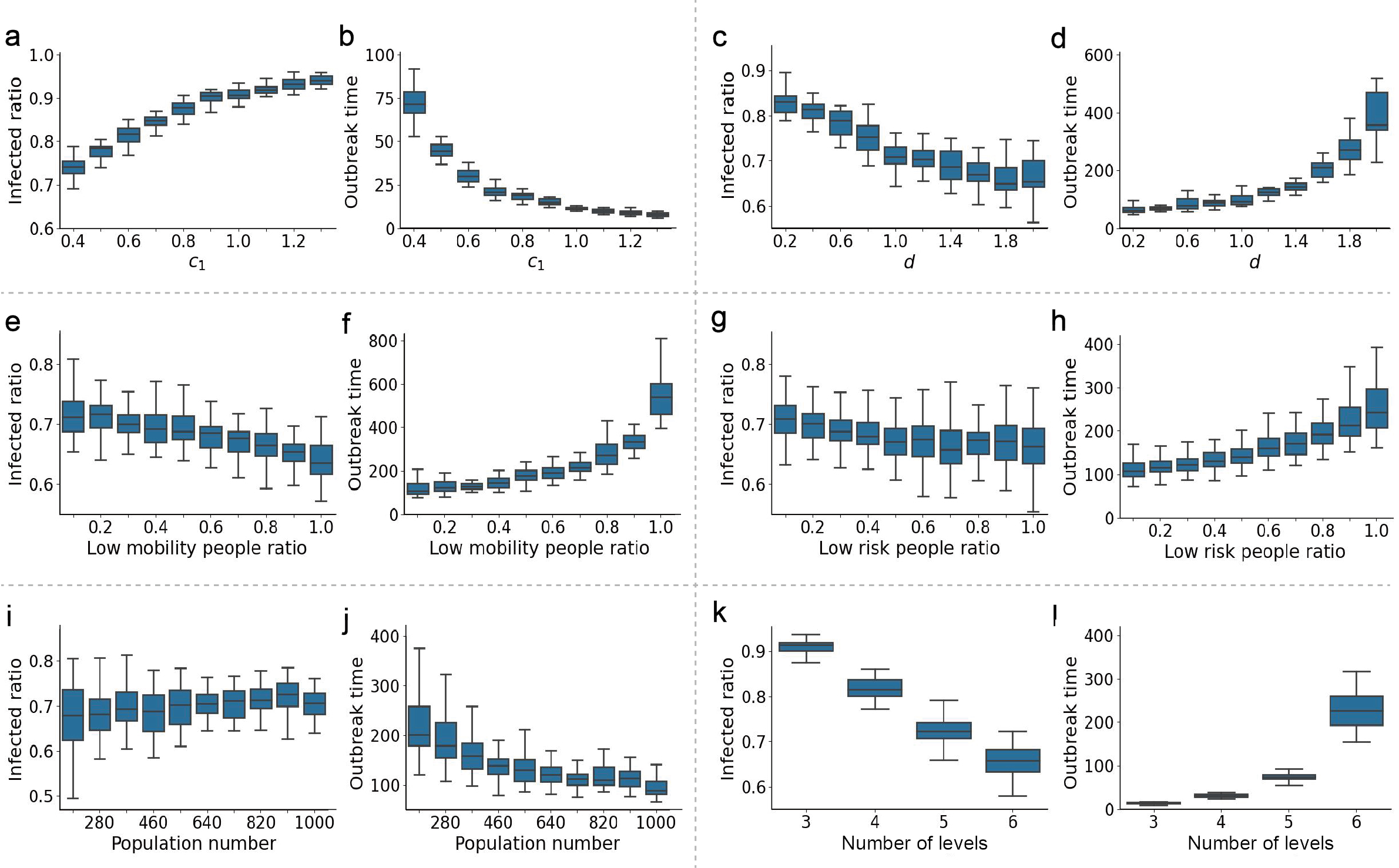}
    \caption{\textbf{The influence of human mobility and health conditions on the disease spread and outbreak.}
    The influence of moving distance regulated by parameter $c_1$ on \textbf{(a)} the infected people ratio at outbreak and \textbf{(b)} the outbreak time. The influence of travelling between levels regulated by parameter $d$ on \textbf{(c)} the infected ratio and \textbf{(d)} the outbreak time. The influence of the percentage of low mobility population on \textbf{(e)} the infected ratio and \textbf{(f)} the outbreak time. The influence of the percentage of low risk population on \textbf{(g)} the infected ratio and \textbf{(h)} the outbreak time. The influence of population size on \textbf{(i)} the infected ratio and \textbf{(j)} the outbreak time. The influence of the number of hierarchical levels on \textbf{(k)} the infected ratio and \textbf{(l)} the outbreak time.}
\end{figure*}

\subsection{Influential factors on disease outbreak}
The spread and outbreak of COVID-19 is influenced by various complex factors, such as biological characteristics of the virus, health and immune conditions, population density, environmental conditions and policies. Using the Mob-Cov model, we mainly decipher the impact of microscopic and universal human mobility and individual health conditions on the disease spread and outbreak (Figure 5). 

For the mobility, the results show that the moving distance and the probability of traveling between large-level containers have significant influence on the disease spread and the onset of the outbreak. When long movements in the same level happen more frequently, which corresponds to a smaller value of $c_1$, the local crowdedness is reduced. Thus, it takes a longer time to reach outbreak (Figure 5b). The infected people are likely to be distributed in every containers and the ratio of the infected population can be relatively small when the outbreak happens (Figure 5a). When individuals are likely to travel between large-level containers (e.g. cities and countries), which corresponds to a small value of $d$, the global transmission of the disease is enhanced. The time it takes for the onset of the outbreak becomes shorter. At the outbreak endpoint, more than 80\% of the population have been infected. Restricting travels between large-level containers helps slow down the spread of the disease and reduce the size of the infected population (Figure 5c and Figure 5d). The ratio of the low mobility population over the total population does not have a significant influence on the infected people number when the outbreak happens (Figure 5e). But if most people prefer to stay at the same place than move, the time it takes for reaching the outbreak endpoint increases from 100 time steps to more than 600 time steps. In addition, the variation of the outbreak time is large when the mobility of most people is low, because the outbreak time now heavily relies on the random initial distribution of the population (Figure 5f).

For the individual health conditions, if a large percentage of the population have low infection risk (e.g. most people get a COVID-19 vaccination), the spread speed of the disease is slowed down (Figure 5h). The low risk people ratio does not have significant influence on the infected population size at the outbreak endpoint. Around 70\% of the people are infected at the outbreak (Figure 5g). The total population size also does not have a significant impact on the infected population ratio (Figure 5i). But when the population number is large, the local crowdedness increases the infection probability of individuals and accelerates the spread of the disease (Figure 5i).

Lastly, we find that the hierarchical structure of the containers has a large impact on the infected population ratio and the outbreak time. If the hidden network of human travels contains many levels, the spread speed of the disease is slower and the outbreak time is larger (Figure 5l). In addition, less people get infected at the outbreak endpoint (Figure 5k). The result suggests that multi-layer geographical barriers and borders that restrict human mobility can help reduce the spread of the disease. For example, in the areas of high population density, the spread of COVID-19 tends to be fast. By zoning and adding multi-level physical borders, the transmission of the disease may be effectively reduced.

\begin{figure*}[!t]
    \centering
    \includegraphics[width=0.95\linewidth]{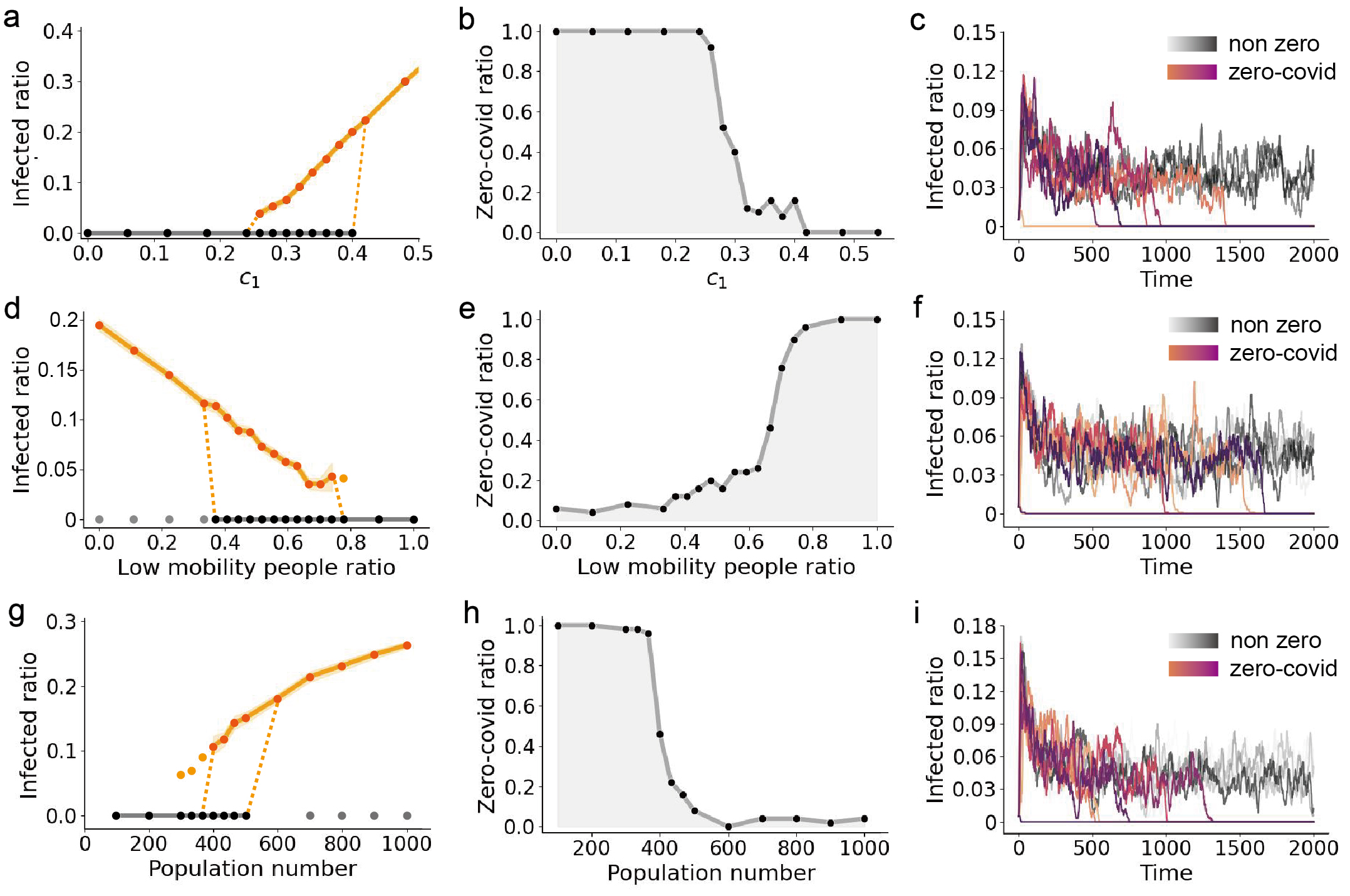}
    \caption{\textbf{Bifurcation diagrams and the transition between "living with COVID" and "zero-COVID" states.} \textbf{(a)} The bifurcation diagram of infected population ratio influenced by mobility parameter $c_1$. Black dots and line are for "zero-COVID" state. Orange dots and line are for "living with COVID" state. Each dot is the mean result of 50 simulations. \textbf{(b)} The percentage of achieving "zero-COVID" influenced by $c_1$. \textbf{(c)} Variation of the infected population ratio over time in the two states when $c_1=0.25$. \textbf{(d)} Bifurcation diagram of infected ratio influenced by the percentage of low mobility people. Grey and light orange dots represent rare cases (1 or 2 out of 50 simulation cases). \textbf{(e)} The percentage of achieving "zero-COVID" influenced by the low mobility people ratio. \textbf{(f)} Variation of the infected ratio over time when the low mobility people ratio is 0.7. \textbf{(g)} Bifurcation diagram of the infected ratio influenced by the population number. \textbf{(h)} The percentage of achieving "zero-COVID" influenced by the population number. \textbf{(i)} Variation of the infected ratio over time when the population number is 400.}
\end{figure*}

\subsection{Bifurcation and state transition}
When the recovery from infection is considered, the system is no longer going to reach the outbreak endpoint. In contrast, depending on the infection rate, recovery speed and rate, and many other factors, two possible states may happen: (1) "zero-COVID" state, which means all the infected people have recovered; (2) "living with COVID" state, during which the system reaching an equilibrium state and the number of newly-infected people is equal to the number of newly-recovered people. In this section, the influential factors on the transition between "zero-COVID" state and "living with COVID" state are investigated.

First, the influence of moving distance determined by parameter $c_1$ is studied. When individuals tend to take long movements in same-level containers (i.e. mobility parameter $c_1$ is smaller than 0.25), the system is able to achieve "zero-COVID". Increasing $c_1$ leads to the local accumulation of people and facilitates the spread of the disease. Bifurcation happens when $c_1$ becomes larger than 0.25. The transition appears to be discontinuous, and a hysteresis loop appears in the bi-stable region where both "zero-COVID" state (black dots) and "living with COVID" state (orange dots) coexist. When $c_1$ is larger than 0.4, the percentage of infected people in the population increases to more than 30\% (Figure 6a). Figure 6b shows that the probability of zero-COVID after a long enough time rapidly decreases to zero when $c_1$ is larger than 0.25. We further show the variation of the infected people ratio over time (Figure 6c). If the system eventually reaches a "zero-COVID" state, the fluctuation curves are labeled in pink colors. Otherwise, the curves are labeled in grey colors. Figure 6c shows that there is a peak of the infected ratio at the beginning, because healthy people keeps being infected but infected people have not recovered yet. Afterwards, the infected ratio decreases and then fluctuates between 3\% to 5\%. In some cases, the infected ratio decreases to zero, indicating that all the sick people become recovered.

Next, effects of the percentage of low mobility people in the population is evaluated. Increasing the percentage of low mobility people reduces the spread of the disease and the infected ratio. Thus, the system transitions from the "living with COVID" state to a bistable region, and then to the "zero-COVID" state (Figure 6d). The probability of achieving zero-COVID increases rapidly when the low mobility people ratio is larger than 60\% (Figure 6e). In addition, the variation of the infected ratio is large and the time for the system to reach the "zero-COVID" state is longer when the low mobility people ratio is 70\% (Figure 6f). This phenomenon may be caused by the stochastic locations and movements of infected people. When the infected people stay in the crowds for a while, it facilitates the disease transmission and increases the infected ratio. But if the infected people are located in a remote area, the low mobility characteristics helps reducing the infection.

Lastly, we also found that varying the population number leads to the bifurcation, and the bistable region with both "zero-COVID" state and "living with COVID" state appears at an intermediate population number (i.e. around 400 to 500 people) (Figure 6g). The probability of zero COVID decreases in the bistable region (Figure 6h). When the population size is larger than 600, the system is likely to reach an equilibrium and has more than 20\% of the population got infected (Figure 6g). When the population number is less than 400, all the infected people can recover from the disease at the first 1000 time steps (Figure 6i) in most simulation cases. 

In addition, effects of infection and recovery probabilities on the disease spread and the infected ratio have also been studied. As expected, increasing the infection probability leads to the increase of infected ratio, while increasing the recovery probability reduces infected ratio. These results are trivial and are not shown here.

\begin{figure*}[!t]
    \centering
    \includegraphics[width=1.0\linewidth]{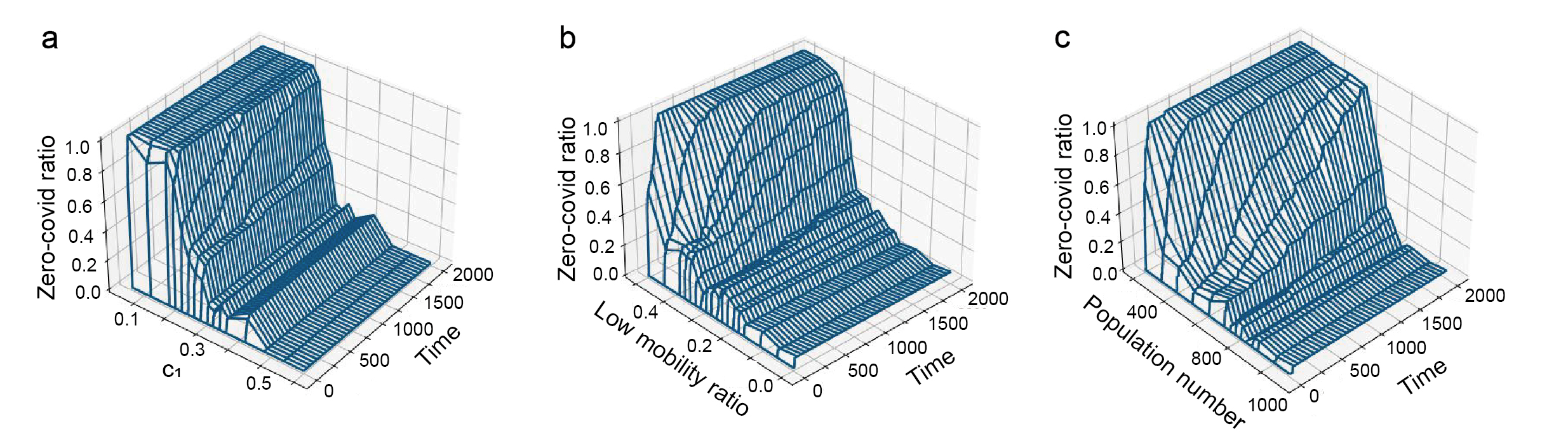}
    \caption{\textbf{Phase diagram of the transient zero-COVID ratio over time influenced by human mobility and population size.} \textbf{(a)} The effect of mobility parameter $c_1$ which represents the traveling distance in the containers. \textbf{(b)} The effect of low mobility people ratio in the population. \textbf{(c)} The effect of the population number.}
\end{figure*}
\subsection{Phase diagram and transient dynamics}
When people take frequent and short local movements, or the population size is large, the spread of the disease is easier and faster. The infected people ratio is larger than 10\%. Reducing local crowdedness helps decrease the infected ratio to around 5\%. Then, fluctuations in the system originated from the random human travel, infection and recovery behaviors may lead to the disappearing of COVID-19 infection (Figures 6c, f and i). The probability to zero COVID-19 and the time it takes are influenced by many factors. In this section, we analyzed the phase diagrams of the infected ratio over time influenced by mobility parameter $c_1$, low mobility people ratio and population number.

Figure 7a shows that the system reaches equilibrium state quickly (i.e. either the "zero-COVID" state or the "living with COVID" state) when $c_1$ is larger than 0.4 or smaller than 0.25. When $c_1 \in [0.25, 0.4]$, the system is bistable and the probability of zero-COVID drops significantly with a small increase of $c_1$. The time for a thorough recovery from COVID-19 becomes longer. From Figure 7b, it can be found that having a large portion of low mobility people in the population (i.e. more than 30\%) increases the time needed for zero-COVID. Though the recovery time is longer, the probability of reaching "zero-COVID" state is higher. When the low mobility people ratio is small, the system is likely to be in the "living with COVID" equilibrium state. Moreover, the population number also influences the transient COVID-19 transmission dynamics (Figure 7c). When the number of people in the population is small (i.e. less than 400), all the people can recover from the disease within the first 500 time steps. When the people number increases, the probability of reaching the zero-COVID state is lower and the population recovery takes a longer time. When the population contains more than 800 people, the newly infected people and recovered people achieves a dynamical equilibrium. The system is in the "live with COVID" state, and more than 20\% of people in the population are infected.

\section{Discussion}
The COVID-19 pandemic has presented an unprecedented challenge and an urgent need for accurate understanding and rapid prediction of the disease dynamics in a wide range of scales. As an airborne disease which transmits mainly through close-contact meetings and interactions, peoples' physical locations and movements are most fundamental sources that drive the spread of COVID-19. However, due to the limited understanding of human mobility, until now, how the intertwined long-distance travels and local movements lead to contagion events and virus spread remains illusive. 

In this study, we developed a mobility-influenced COVID-19 transmission model (Mob-Cov) using the agent-based Monte Carlo modeling strategy and the multi-level nested spatial containers. Human mobility behavior is restricted based on a hierarchical structure of spatial containers, corresponding to the geographical places from buildings, via streets and cities, to nations and continents. By merging the power-law movements within containers and the macroscopic transport between containers, the generated human travel patterns are highly realistic, which ensures an accurate study of the transmission dynamics of COVID-19 influenced by mobility. In addition, due to the flexibility of agent-based modeling strategy, different individual moving actions and health conditions are considered in the model. We found that long travels within a small-level container (e.g. a street or a county) helps alleviate local crowdedness of people and reduces infection. But traveling between levels and large-level containers (e.g. traveling between cities and nations) converts the local transmission to global disease spread, facilitating the outbreak of the disease. Moreover, the infection and recovery of an individual leads to the bifurcation and the appearance of two dynamic states, "zero-COVID" state and "live with COVID" state. We conducted bifurcation analysis and found that the ability of zero COVID, which means all the infected people are recovered, is largely influenced by human mobility and population number. The Mob-Cov model might be further used to quantitatively measure how much local accumulation of people and global travels need to be reduced in order to achieve successfully removing the COVID-19 infection from the population. 

Overall, the Mob-Cov model can be extended in four major directions in future works. (1) Infection in a wide range of spatial scales can be considered. Specifically, compared with previous works, infection on a contact network estimated from human GPS data mainly captures the transmission between cities and nations\cite{21,22,23}. Agent-based models using random walks to describe human movements can only study the microscopic actions within a building or a district\cite{16,17,18,19}. The multi-level hierarchical container structures in the Mob-Cov provides a more realistic and accurate description of individuals moving cross multiple length scales while maintaining a reasonable computational cost. (2) The nested container structures of human mobility can be estimated from the real-world human GPS data using maximum likelihood estimator\cite{24}. Thus, the model is capable of predicting the distribution of COVID-19 infection in reality and providing the early warning of disease outbreak based on the past infection data and the geographical information. (3) Various interventions can also be easily implemented and used to study the impact of policies on the control of the COVID-19 pandemic, especially the effective length scales of these policies. (4) In this work, we focused on modeling the multi-level human mobility while implementing only simple rules of disease infection. It would be interesting to further extend the Mob-Cov model to complex infection scenarios, including the susceptible, exposed, infectious and recovered states. In particular, the susceptible and exposed states can be critical, since the latent period between being infected and becoming infectious has made the disease more contagious. Moreover, whether the epidemic wave changes on the nested mobility networks, and how does it be influenced by population characteristics (e.g. age, gender, level of urbanization) should be further investigated using our Mob-Cov model.

\section*{Acknowledgments}
We acknowledge support from the National Natural Science Foundation of China (Grant No.~12102081) and Fundamental Research Funds for the Central Universities in China (Grant No.~DUT21RC(3)044).

\section*{Data Availability}
The data that support the findings of this study are available from the corresponding author upon reasonable request.

\section*{Conflict of interest}
The authors declare that they have no conflict of interest.

\bibliographystyle{IEEEtran}

\bibliography{sn-bibliography}

\begin{thebibliography}{10}
\providecommand{\url}[1]{#1}
\csname url@samestyle\endcsname
\providecommand{\newblock}{\relax}
\providecommand{\bibinfo}[2]{#2}
\providecommand{\BIBentrySTDinterwordspacing}{\spaceskip=0pt\relax}
\providecommand{\BIBentryALTinterwordstretchfactor}{4}
\providecommand{\BIBentryALTinterwordspacing}{\spaceskip=\fontdimen2\font plus
\BIBentryALTinterwordstretchfactor\fontdimen3\font minus
  \fontdimen4\font\relax}
\providecommand{\BIBforeignlanguage}[2]{{%
\expandafter\ifx\csname l@#1\endcsname\relax
\typeout{** WARNING: IEEEtran.bst: No hyphenation pattern has been}%
\typeout{** loaded for the language `#1'. Using the pattern for}%
\typeout{** the default language instead.}%
\else
\language=\csname l@#1\endcsname
\fi
#2}}
\providecommand{\BIBdecl}{\relax}
\BIBdecl

\bibitem{1}
T.~P. Velavan and C.~G. Meyer, ``The covid‐19 epidemic,'' \emph{Trop Med Int
  Health}, vol.~25, pp. 278--280, 2020.

\bibitem{2}
P.~K. Ozili and T.~Arun, ``Spillover of covid-19: impact on the global
  economy,'' in \emph{Managing Inflation and Supply Chain Disruptions in the
  Global Economy}.\hskip 1em plus 0.5em minus 0.4em\relax IGI Global, 2023, pp.
  41--61.

\bibitem{3}
O.~J. Watson, G.~Barnsley, J.~Toor, A.~B. Hogan, P.~Winskill, and A.~C. Ghani,
  ``Global impact of the first year of covid-19 vaccination: a mathematical
  modelling study,'' \emph{The Lancet Infectious Diseases}, 2022.

\bibitem{4}
O.~N. Bj{\o}rnstad, K.~Shea, M.~Krzywinski, and N.~Altman, ``The seirs model
  for infectious disease dynamics.'' \emph{Nature methods}, vol.~17, no.~6, pp.
  557--559, 2020.

\bibitem{5}
F.~Nda{\"\i}rou, I.~Area, J.~J. Nieto, and D.~F. Torres, ``Mathematical
  modeling of covid-19 transmission dynamics with a case study of wuhan,''
  \emph{Chaos, Solitons \& Fractals}, vol. 135, p. 109846, 2020.

\bibitem{6}
J.~Dehning, J.~Zierenberg, F.~P. Spitzner, M.~Wibral, J.~P. Neto, M.~Wilczek,
  and V.~Priesemann, ``Inferring change points in the spread of covid-19
  reveals the effectiveness of interventions,'' \emph{Science}, vol. 369, no.
  6500, p. eabb9789, 2020.

\bibitem{7}
G.~Giordano, F.~Blanchini, R.~Bruno, P.~Colaneri, A.~Di~Filippo, A.~Di~Matteo,
  and M.~Colaneri, ``Modelling the covid-19 epidemic and implementation of
  population-wide interventions in italy,'' \emph{Nature medicine}, vol.~26,
  no.~6, pp. 855--860, 2020.

\bibitem{8}
Q.~Jia, J.~Li, H.~Lin, F.~Tian, and G.~Zhu, ``The spatiotemporal transmission
  dynamics of covid-19 among multiple regions: a modeling study in chinese
  provinces,'' \emph{Nonlinear Dynamics}, vol. 107, no.~1, pp. 1313–--1327,
  2022.

\bibitem{9}
Z.~Ma, S.~Wang, X.~Lin, X.~Li, X.~Han, H.~Wang, and H.~Liu, ``Modeling for
  covid-19 with the contacting distance,'' \emph{Nonlinear Dynamics}, vol. 107,
  no.~1, p. 3065–3084, 2022.

\bibitem{9extra}
R.~Ikram, A.~Khan, M.~Zahri, A.~Saeed, M.~Yavuz, and P.~Kumam, ``Extinction and
  stationary distribution of a stochastic covid-19 epidemic model with
  time-delay,'' \emph{Computers in Biology and Medicine}, vol. 141, p. 105115,
  2022.

\bibitem{9extra2}
S.~G. Babajanyan and K.~H. Cheong, ``Age-structured sir model and resource
  growth dynamics: a covid-19 study,'' \emph{Nonlinear Dynamics}, vol. 104, p.
  2853–2864, 2021.

\bibitem{10}
S.~L. Chang, N.~Harding, C.~Zachreson, O.~M. Cliff, and M.~Prokopenko,
  ``Modelling transmission and control of the covid-19 pandemic in australia,''
  \emph{Nature communications}, vol.~11, no.~1, pp. 1--13, 2020.

\bibitem{11}
J.~R. Koo, A.~R. Cook, M.~Park, Y.~Sun, H.~Sun, J.~T. Lim, C.~Tam, and B.~L.
  Dickens, ``Interventions to mitigate early spread of sars-cov-2 in singapore:
  a modelling study,'' \emph{The Lancet Infectious Diseases}, vol.~20, no.~6,
  pp. 678--688, 2020.

\bibitem{12}
N.~Hoertel, M.~Blachier, C.~Blanco, M.~Olfson, M.~Massetti, M.~S. Rico,
  F.~Limosin, and H.~Leleu, ``A stochastic agent-based model of the sars-cov-2
  epidemic in france,'' \emph{Nature medicine}, vol.~26, no.~9, pp. 1417--1421,
  2020.

\bibitem{13}
M.~Shamil, F.~Farheen, N.~Ibtehaz, I.~M. Khan, M.~S. Rahman \emph{et~al.}, ``An
  agent-based modeling of covid-19: validation, analysis, and
  recommendations,'' \emph{Cognitive Computation}, pp. 1--12, 2021.

\bibitem{14}
R.~Hinch, W.~J. Probert, A.~Nurtay, M.~Kendall, C.~Wymant, M.~Hall, K.~Lythgoe,
  A.~Bulas~Cruz, L.~Zhao, A.~Stewart \emph{et~al.}, ``Openabm-covid19—an
  agent-based model for non-pharmaceutical interventions against covid-19
  including contact tracing,'' \emph{PLoS computational biology}, vol.~17,
  no.~7, p. e1009146, 2021.

\bibitem{15}
C.~C. Kerr, R.~M. Stuart, D.~Mistry, R.~G. Abeysuriya, K.~Rosenfeld, G.~R.
  Hart, R.~C. N{\'u}{\~n}ez, J.~A. Cohen, P.~Selvaraj, B.~Hagedorn
  \emph{et~al.}, ``Covasim: an agent-based model of covid-19 dynamics and
  interventions,'' \emph{PLOS Computational Biology}, vol.~17, no.~7, p.
  e1009149, 2021.

\bibitem{16}
E.~Cuevas, ``An agent-based model to evaluate the covid-19 transmission risks
  in facilities,'' \emph{Computers in biology and medicine}, vol. 121, p.
  103827, 2020.

\bibitem{17}
P.~C. Silva, P.~V. Batista, H.~S. Lima, M.~A. Alves, F.~G. Guimar{\~a}es, and
  R.~C. Silva, ``Covid-abs: An agent-based model of covid-19 epidemic to
  simulate health and economic effects of social distancing interventions,''
  \emph{Chaos, Solitons \& Fractals}, vol. 139, p. 110088, 2020.

\bibitem{17extra}
G.~A. Palomo-Briones, M.~Siller, and A.~Grignard, ``An agent-based model of the
  dual causality between individual and collective behaviors in an epidemic,''
  \emph{Computers in Biology and Medicine}, vol. 141, p. 104995, 2022.

\bibitem{18}
A.~Rodr{\'\i}guez, E.~Cuevas, D.~Zaldivar, B.~Morales-Casta{\~n}eda, R.~Sarkar,
  and E.~H. Houssein, ``An agent-based transmission model of covid-19 for
  re-opening policy design,'' \emph{Computers in Biology and Medicine}, vol.
  148, p. 105847, 2022.

\bibitem{19}
T.~Kano, K.~Yasui, T.~Mikami, M.~Asally, and A.~Ishiguro, ``An agent-based
  model of the interrelation between the covid-19 outbreak and economic
  activities,'' \emph{Proceedings of the Royal Society A}, vol. 477, no. 2245,
  p. 20200604, 2021.

\bibitem{19extra}
G.~Lombardo, M.~Pellegrino, M.~Tomaiuolo, S.~Cagnoni, M.~Mordonini,
  M.~Giacobini, and A.~Poggi, ``Fine-grained agent-based modeling to predict
  covid-19 spreading and effect of policies in large-scale scenarios,''
  \emph{IEEE Journal of Biomedical and Health Informatics}, vol.~26, no.~5, pp.
  2052--2062, 2022.

\bibitem{19extra2}
P.~Ciunkiewicz, W.~Brooke, M.~Rogers, and S.~Yanushkevich, ``Agent-based
  epidemiological modeling of covid-19 in localized environments,''
  \emph{Computers in Biology and Medicine}, vol. 144, p. 105396, 2022.

\bibitem{19extra3}
S.~Yin and N.~Zhang, ``Prevention schemes for future pandemic cases:
  mathematical model and experience of interurban multi-agent covid-19 epidemic
  prevention,'' \emph{Nonlinear Dynamics}, vol. 104, p. 2865–2900, 2021.

\bibitem{20}
L.~Alessandretti, ``What human mobility data tell us about covid-19 spread,''
  \emph{Nature Reviews Physics}, vol.~4, no.~1, pp. 12--13, 2022.

\bibitem{21}
N.~E. Kogan, L.~Clemente, P.~Liautaud, J.~Kaashoek, N.~B. Link, A.~T. Nguyen,
  F.~S. Lu, P.~Huybers, B.~Resch, C.~Havas \emph{et~al.}, ``An early warning
  approach to monitor covid-19 activity with multiple digital traces in near
  real time,'' \emph{Science Advances}, vol.~7, no.~10, p. eabd6989, 2021.

\bibitem{22}
A.~Aleta, D.~Martin-Corral, A.~Pastore~y Piontti, M.~Ajelli, M.~Litvinova,
  M.~Chinazzi, N.~E. Dean, M.~E. Halloran, I.~M. Longini~Jr, S.~Merler
  \emph{et~al.}, ``Modelling the impact of testing, contact tracing and
  household quarantine on second waves of covid-19,'' \emph{Nature Human
  Behaviour}, vol.~4, no.~9, pp. 964--971, 2020.

\bibitem{23}
D.~Mistry, M.~Litvinova, A.~Pastore~y Piontti, M.~Chinazzi, L.~Fumanelli, M.~F.
  Gomes, S.~A. Haque, Q.-H. Liu, K.~Mu, X.~Xiong \emph{et~al.}, ``Inferring
  high-resolution human mixing patterns for disease modeling,'' \emph{Nature
  communications}, vol.~12, no.~1, pp. 1--12, 2021.

\bibitem{24extra1}
F.~Xu, Y.~Li, D.~Jin, J.~Lu, and C.~Song, ``Emergence of urban growth patterns
  from human mobility behavior,'' \emph{Nature Computational Science}, vol.~1,
  no.~12, pp. 791--800, 2021.

\bibitem{24extra2}
C.~Song, T.~Koren, P.~Wang, and A.-L. Barab{\'a}si, ``Modelling the scaling
  properties of human mobility,'' \emph{Nature physics}, vol.~6, no.~10, pp.
  818--823, 2010.

\bibitem{24extra3}
G.~Solmaz and D.~Turgut, ``A survey of human mobility models,'' \emph{IEEE
  Access}, vol.~7, pp. 125\,711--125\,731, 2019.

\bibitem{24}
L.~Alessandretti, U.~Aslak, and S.~Lehmann, ``The scales of human mobility,''
  \emph{Nature}, vol. 587, no. 7834, pp. 402--407, 2020.

\bibitem{25}
C.~Song, T.~Koren, P.~Wang, and A.-L. Barab{\'a}si, ``Modelling the scaling
  properties of human mobility,'' \emph{Nature physics}, vol.~6, no.~10, pp.
  818--823, 2010.

\bibitem{28}
C.~Song, T.~Koren, P.~Wang, and A.-L. Barab\'asi, ``Modelling the scaling
  properties of human mobility,'' \emph{Nature physics}, vol.~6, no.~10, pp.
  818--823, 2010.

\bibitem{29}
F.~Xu, Y.~Li, D.~Jin, J.~Lu, and C.~Song, ``Emergence of urban growth patterns
  from human mobility behavior,'' \emph{Nature Computational Science}, vol.~1,
  no.~12, pp. 791--800, 2021.

\bibitem{29extra1}
P.~Deville, C.~Song, N.~Eagle, V.~D. Blondel, A.-L. Barab{\'a}si, and D.~Wang,
  ``Scaling identity connects human mobility and social interactions,''
  \emph{Proceedings of the National Academy of Sciences}, vol. 113, no.~26, pp.
  7047--7052, 2016.

\bibitem{29extra2}
L.~Pappalardo, S.~Rinzivillo, and F.~Simini, ``Human mobility modelling:
  exploration and preferential return meet the gravity model,'' \emph{Procedia
  Computer Science}, vol.~83, pp. 934--939, 2016.

\bibitem{30}
V.~Satopaa, J.~Albrecht, D.~Irwin, and B.~Raghavan, ``Finding a "kneedle" in a
  haystack: Detecting knee points in system behavior,'' in \emph{2011 31st
  international conference on distributed computing systems workshops}.\hskip
  1em plus 0.5em minus 0.4em\relax IEEE, 2011, pp. 166--171.

\end{thebibliography}


\end{document}